# Longitudinal wearable monitoring and polygenic risk for incident major depressive disorder in the All of Us Research Program

Yuezhou Zhang[1,*], Amos A. Folarin[1,2,3,4,5], Rongrong Zhong[1,6], Hyunju Kim[1], Shaoxiong Sun[1,7], Callum Stewart[1], Richard JB Dobson[1,2,3,4,5,*]

[1]Department of Biostatistics & Health Informatics, Institute of Psychiatry, Psychology and Neuroscience, King's College London, London, United Kingdom

[2]Institute of Health Informatics, University College London, London, United Kingdom

[3]NIHR Biomedical Research Centre at South London and Maudsley, NHS Foundation Trust, London, United Kingdom

[4]NIHR Biomedical Research Centre at University College London Hospitals, NHS Foundation Trust, London, United Kingdom

[5]Health Data Research UK, University College London, London, United Kingdom

[6]Clinical Research Center & Division of Mood Disorders, Shanghai Mental Health Center, Shanghai Jiao Tong University School of Medicine, Shanghai, China

[7]Department of Computer Science, University of Sheffield, Sheffield, United Kingdom

[*]Corresponding authors: Yuezhou Zhang (yuezhou.zhang@kcl.ac.uk) and Richard JB Dobson (richard.j.dobson@kcl.ac.uk)

## Abstract

Major depressive disorder (MDD) risk reflects both stable inherited liability and dynamic behavioral patterns, yet these dimensions are rarely examined together using long-term objective data in real-world settings. Here, we integrated genomic, electronic health record (EHR), and longitudinal Fitbit wearable data from 3,030 adults of genetically inferred European ancestry in the All of Us Research Program, 284 of whom developed EHR-recorded incident MDD after a 180-day baseline period. Time-varying Cox models examined associations of MDD polygenic risk scores (PRS), monthly wearable-derived physical activity and sleep features, and wearable feature × MDD PRS interactions with incident MDD. Higher MDD PRS, lower daily steps, lower light and vigorous physical activity, lower sleep efficiency, and greater sleep duration variability were associated with higher risk of EHR-recorded incident MDD. The associations of sedentary time and sleep duration variability with incident MDD differed across MDD PRS levels, with stronger risk associations among participants with higher MDD PRS. PRS-stratified hazard ratio curves further indicated that comparable estimated risk corresponded to more favorable behavioral levels (such as higher daily step counts and more stable sleep) among participants with higher MDD PRS than among those with lower MDD PRS. Sequentially integrating MDD PRS, baseline wearable features, monthly wearable features, and selected wearable feature × MDD PRS interactions increased model discrimination, with the C-index increasing from 0.637 to 0.705. These findings support the complementary value of inherited liability and longitudinal real-world behavioral monitoring for incident MDD risk characterization and may inform future work on genetically informed digital phenotyping for personalized risk monitoring and prevention.

# Introduction

Major depressive disorder (MDD) is a prevalent and disabling mental disorder worldwide, imposing substantial individual and societal burden [1]. Identifying individuals at elevated MDD risk may inform preventive strategies and support timely intervention [2, 3]. Yet this remains challenging because MDD risk reflects both inherited liability [4, 5] and dynamic behavioral patterns (e.g., physical activity and sleep) that may contribute to risk or signal emerging changes in mental health [6-8]. These distinct dimensions of risk are difficult to capture through conventional clinical assessments, which are typically episodic and often rely on retrospective self-report and clinical expertise [9]. Longitudinal mobile monitoring (e.g., wearables and smartphones) can objectively capture real-world behavioral patterns [10], whereas genetic risk measures can quantify a stable component of inherited liability [11]. Integrating these complementary measures may therefore improve the characterization of MDD risk in daily life.

Genetic studies indicate that inherited factors contribute substantially to MDD risk, with population-level heritability estimates of approximately 30% to 40% [4, 5]. Large-scale genome-wide association studies (GWAS) have further shown that this liability is highly polygenic, involving many common variants with individually small effects [12, 13]. Polygenic risk scores (PRS) aggregate genome-wide genetic effects into an individual-level measure of inherited liability and have been increasingly used for MDD risk stratification [14, 15]. However, PRS captures only the inherited component of liability and explains a modest proportion of variation in overall MDD risk [16]. Risk characterization may therefore benefit from combining MDD PRS with dynamic, non-genetic phenotypes that capture behavioral patterns relevant to mental health [17].

Physical activity and sleep are behavioral phenotypes with established relevance to depression. Meta-analyses and prospective cohort studies have identified low physical activity as a risk factor for incident depression [6, 18, 19], whereas higher physical activity has been associated with a lower risk of depression onset, consistent with a potentially protective role [20, 21]. Sleep disturbances and irregular sleep patterns have likewise been associated with subsequent depression risk [22-24]. However, much of this evidence relies on cross-sectional or intermittently self-reported behaviors, which are susceptible to recall bias and may not capture dynamic, longitudinal changes in daily-life behavior. Wearable sensing technologies provide a low-burden approach to measuring physical activity and sleep objectively in real-world settings, with growing evidence linking wearable-derived physical activity and sleep features to depressive symptom severity and depression risk [8, 25-27]. Because genetic liability may influence susceptibility to behavioral exposures or the extent to which behavioral changes reflect emerging vulnerability, the same wearable-derived phenotype may carry different risk information across individuals.

These considerations point to a central evidence gap: long-term wearable-derived behavioral phenotypes have rarely been integrated with genetic liability for MDD risk characterization. Prior work has shown that higher self-reported physical activity is associated with lower incident depression risk across all genetic liability levels, but physical activity was assessed using a single questionnaire-based measure [28]. Another study linked higher depression PRS to lower baseline physical activity measured over a 7-day period, but did not examine incident MDD risk or longitudinal behavioral patterns [29]. More recently, a preprint involving adolescents from the Adolescent Brain Cognitive Development (ABCD) study used approximately 3 weeks of Fitbit data and reported stronger associations between Fitbit-derived behavioral features and depressive symptom severity at higher levels of depression PRS [30]. However, this evidence focused on adolescents rather than adults and did not examine longer-term wearable data. Although integrating wearable-derived behaviors with PRS has begun to support risk stratification for incident cardiometabolic and cardiovascular outcomes, including obesity, type 2 diabetes, and atrial fibrillation [31-33], comparable approaches remain limited for MDD. It remains unclear whether long-term wearable-derived behavioral phenotypes can complement MDD PRS for incident risk characterization, and whether these digital phenotypes carry different risk information across levels of genetic liability.

To address this gap, this study aimed to characterize incident MDD recorded in electronic health records (EHR) by integrating MDD PRS with longitudinal wearable-derived physical activity and sleep features in the All of Us Research Program (AoURP) [34]. We further examined whether associations between wearable-derived behavioral features and incident MDD differed across levels of genetic liability, and whether comparable estimated risk corresponded to different behavioral levels across PRS strata. Finally, we evaluated the incremental value of sequentially integrating multiple sources of information for MDD risk characterization. By linking stable inherited liability with dynamic real-world behavioral monitoring and EHR-recorded diagnoses, this study evaluated a genetically informed digital phenotyping framework for MDD.

## Results

### Analytic cohort

The overall study design and analytic framework are shown in Figure 1. The analytic cohort included 3,030 participants of genetically inferred European ancestry with linked genomic, EHR, and Fitbit data and no EHR-recorded MDD diagnosis before Fitbit monitoring or during the 180-day baseline wearable period. Of these participants, 284 developed EHR-recorded incident MDD after the baseline wearable period. The cohort construction flow chart is shown in Supplementary Figure 1. The median age at enrollment was 52.4 years (interquartile range [IQR], 36.9–62.7), 2,113 of 2,959 participants with available sex information were female (71.4%), and the median body mass index (BMI) was 27.6 kg/m$^2$ (IQR, 24.2–32.2). Participants had a median Fitbit monitoring duration of 5.05 years (IQR, 3.08–6.62). Participant characteristics and baseline wearable features are summarized in Table 1, overall and stratified by MDD PRS quartile. Participants in the highest MDD PRS quartile had the highest median BMI, the lowest proportion with a college degree, the highest prevalence of lifetime smoking, the lowest median baseline daily steps, and the highest baseline sleep duration variability.

**MDD PRS performance in the broader EHR sample**

In the analysis separate from the primary Fitbit-linked analytic cohort, we assessed the performance of the calculated MDD PRS in a broader sample of 110,872 participants of genetically inferred European ancestry who had EHR data and genomic data from which MDD PRS could be derived. In logistic regression adjusted for age and sex, each standard deviation increase in MDD PRS was associated with 36% higher odds of any recorded MDD diagnosis (odds ratio [OR] = 1.36; $P < 0.001$).

**Associations of MDD PRS and time-varying wearable-derived features with EHR-recorded incident MDD**

We first examined whether MDD PRS and monthly time-varying wearable-derived features were associated with EHR-recorded incident MDD. Full estimates of all time-varying Cox models are provided in Supplementary Tables 3 and 4. Across the separate wearable feature models, higher MDD PRS was consistently associated with greater incident MDD risk, with hazard ratios (HRs) for the 75th versus 25th percentile contrast ranging from 1.23 to 1.38 ($P <0.001$).

Several time-varying wearable-derived features were significantly associated with incident MDD risk (Table 2). Among physical activity features, higher daily steps (HR = 0.59 for 10.87 vs 5.95 thousand steps/day; $P < 0.001$), light physical activity (HR = 0.66 for 274.33 vs 186.24 min/day; $P < 0.001$), and vigorous physical activity (HR = 0.78 for 34.42 vs 6.31 min/day; $P = 0.027$) were associated with lower incident MDD risk. Among sleep features, greater sleep duration variability was associated with higher risk (HR = 1.29 for 1.56 vs 0.90 h; $P < 0.001$), whereas higher sleep efficiency was associated with lower risk (HR = 0.90 for 92% vs 87%; $P = 0.039$).

Sleep duration was the only wearable-derived feature showing evidence of nonlinearity. Based on the restricted cubic spline model, the model-estimated lowest-risk sleep duration was 6.71 h. Compared with this value, shorter sleep durations were associated with higher

MDD risk, particularly at 4 h (HR = 1.69, P = 0.005) and 5 h (HR = 1.35, P = 0.007), whereas longer sleep durations showed numerically higher but non-significant risk estimates (Supplementary Table 2).

**PRS-stratified wearable associations with EHR-recorded incident MDD**

We next examined whether associations between wearable-derived behavioral features and incident MDD risk differed across MDD PRS levels by including wearable-derived feature × MDD PRS interaction terms in the time-varying Cox models. Two features showed statistically significant interactions with MDD PRS: sedentary time and sleep duration variability (Table 3). Figure 2 shows how the associations between selected wearable features and incident MDD varied across MDD PRS levels, showing model-estimated HR curves at the 25th, 50th, and 75th percentiles of MDD PRS.

For sedentary time, higher levels were associated with greater incident MDD risk only at higher MDD PRS. At the 75th percentile of MDD PRS, sedentary time at the 75th versus 25th percentile was associated with higher incident MDD risk (HR = 1.22; P = 0.009), whereas no significant association was observed at the 25th percentile (HR = 0.98; P = 0.822) or 50th percentile of MDD PRS (HR = 1.09; P = 0.225). Consistent with this interaction, higher sedentary time was associated with little change in model-estimated HRs at the 25th and 50th percentiles of MDD PRS, but with progressively higher HRs at the 75th percentile of MDD PRS (Figure 2b).

For sleep duration variability, the association with incident MDD strengthened across increasing MDD PRS percentiles. For the 75th versus 25th percentile contrast in sleep duration variability, the HR increased from 1.18 at the 25th percentile of MDD PRS (P = 0.077) to 1.29 at the 50th percentile (P < 0.001) and 1.42 at the 75th percentile (P < 0.001). The PRS-stratified curves showed steeper positive gradients at higher MDD PRS levels (Figure 2c), indicating a stronger association between greater sleep duration variability and incident MDD risk among participants with higher MDD PRS. No other wearable feature × MDD PRS interaction terms were statistically significant.

**PRS-stratified behavioral levels at comparable estimated risk**

We further examined whether comparable estimated risk of incident MDD corresponded to different behavioral levels across MDD PRS levels. For each wearable feature, we identified the feature value at which each PRS-stratified HR curve crossed HR = 1.00, corresponding to no higher or lower model-estimated risk on the HR scale.

For daily steps (Figure 2a), the feature level corresponding to HR = 1.00 increased across MDD PRS percentiles, from 6,768 steps/day at the 25th percentile to 8,208 steps/day at the 50th percentile and 9,842 steps/day at the 75th percentile. For sedentary time (Figure 2b), the HR = 1.00 level was not estimable at the 25th percentile of MDD PRS; among the estimable curves, the corresponding sedentary time level was lower at the 75th percentile than at the 50th percentile of MDD PRS (615.4 vs 701.4 min/day). For sleep duration variability (Figure 2c), the feature level corresponding to HR = 1.00 decreased across MDD PRS percentiles, from 1.60 h at the 25th percentile to 1.19 h at the 50th percentile and 0.99 h at the 75th percentile.

Sleep duration showed similar U-shaped HR curves across MDD PRS levels (Figure 2d). Although risk was lowest near the model-estimated lowest-risk sleep duration within each PRS level, the curves for the 50th and 75th percentiles of MDD PRS remained above HR = 1.00 across the evaluated range. Thus, a more favorable sleep duration was associated with lower estimated risk, but the corresponding HR did not reach 1.00 at higher MDD PRS.

Overall, comparable estimated risk generally corresponded to more favorable behavioral levels among participants with higher MDD PRS, including higher daily steps, less sedentary time, lower sleep duration variability, greater light and vigorous physical activity, and higher sleep efficiency; moderate physical activity showed no clear pattern. Results for the remaining wearable features are shown in Supplementary Figure 2.

**Incremental contribution to incident MDD risk characterization**

To assess the incremental contribution of each information domain to incident MDD risk characterization, we sequentially added base covariates, MDD PRS, the full set of baseline wearable features, the full set of monthly time-varying wearable features, and statistically significant wearable feature × MDD PRS interaction terms from the primary Cox models. Each model extension showed a statistically significant improvement in model fit compared with the preceding nested model based on the likelihood ratio test (Table 4). Model performance was summarized using Akaike information criterion (AIC), where lower values indicate better model fit, and Harrell's C-index, where higher values indicate better discrimination.

From Model 1 (the base covariate model) to Model 5 (the final integrated model), C-index increased from 0.637 to 0.705, while AIC decreased from 4222.66 to 4156.18. Adding MDD PRS to the base covariate model improved model fit and discrimination (ΔAIC = −13.35; ΔC-index = 0.013). Further improvements were observed after adding baseline wearable features (ΔAIC = −25.98; ΔC-index = 0.029) and monthly time-varying wearable features (ΔAIC = −21.18; ΔC-index = 0.024). Adding the selected sedentary time × MDD PRS and sleep duration variability × MDD PRS interaction terms provided a smaller but statistically significant improvement in model fit (ΔAIC = −5.98; ΔC-index = 0.002).

**Sensitivity analysis**

To account for potential reverse causation and diagnostic delay, we repeated the primary Cox analyses after excluding wearable observations within 6 months before the first EHR-recorded incident MDD diagnosis (Supplementary Table 5). The overall pattern of associations between monthly wearable features and incident MDD was broadly consistent with the primary analyses, although the association for vigorous physical activity was no longer statistically significant. The sleep duration variability × MDD PRS and sedentary time × MDD PRS interactions remained statistically significant and in the same direction as the primary analyses, supporting the robustness of the PRS-stratified associations.

## Discussion

In this study, we integrated MDD PRS with long-term wearable-derived physical activity and sleep features to characterize EHR-recorded incident MDD risk within a longitudinal framework. Higher MDD PRS and less favorable longitudinal behavioral patterns, including lower physical activity, greater sleep variability, and lower sleep efficiency, were associated with higher incident MDD risk. Importantly, the associations of sedentary time and sleep duration variability with incident MDD differed across MDD PRS levels, suggesting that these wearable-derived behavioral phenotypes carried different risk information across levels of genetic liability. In addition, comparable estimated risk corresponded to more favorable behavioral levels among participants with higher MDD PRS. Sequential model comparisons further showed that MDD PRS, baseline wearable features, monthly time-varying wearable features, and PRS-stratified behavioral associations each provided incremental information for MDD risk characterization. Together, these findings support a genetically informed digital phenotyping framework that links stable inherited liability with dynamic, real-world behavioral monitoring.

A key finding was that the strength of the associations of sedentary time and sleep duration variability with EHR-recorded incident MDD differed across MDD PRS levels, as indicated by statistically significant wearable feature × MDD PRS interaction terms. The association with sedentary time became more positive at higher MDD PRS, whereas the association with sleep duration variability strengthened progressively across PRS percentiles. Psychiatric genetics research has suggested that susceptibility to stress-related exposures may differ across levels of genetic liability [35], and prior work has also shown that individuals with both high depression genetic risk and unfavorable lifestyle profiles have the highest risk of incident depression [17]. In addition, a recent preprint involving adolescents from the ABCD study similarly reported stronger associations between Fitbit-derived behavioral features and depressive symptom severity at higher levels of depression PRS, although the study population differed from our adult cohort [30]. Our findings suggest that the risk information carried by wearable-derived behavioral phenotypes may differ across levels of genetic liability, with sedentary behavior and irregular sleep potentially representing particularly informative risk phenotypes among individuals with higher MDD PRS.

As a complementary analysis, we identified the wearable-derived behavioral levels corresponding to neither higher nor lower model-estimated risk within each MDD PRS level (HR = 1.00). Participants with higher MDD PRS generally had more favorable behavioral levels at this neutral point, including higher daily steps and physical activity, less sedentary time, and lower sleep duration variability. Given prior evidence that low physical activity is a potentially modifiable factor relevant to depression risk [6, 36], these findings raise the possibility that behavioral levels, and potentially the degree of behavioral change relevant to risk reduction, may differ across levels of genetic liability. They may therefore inform future work on genetically informed behavioral risk stratification and prevention, while recognizing that these model-derived estimates are exploratory, require validation in intervention studies, and should not be interpreted as causal thresholds or clinical intervention recommendations.

The associations of monthly wearable-derived behavioral features with incident MDD were broadly consistent with prior AoURP studies linking Fitbit-derived step count and sleep patterns to chronic disease incidence, including MDD [25, 26]. Our study extends this work by incorporating MDD PRS, broadening physical activity assessment beyond step count to sedentary time and intensity-specific physical activity, and evaluating PRS-stratified associations. Lower physical activity may reflect reduced energy, motivation, or behavioral engagement, whereas irregular and inefficient sleep may reflect disrupted sleep–wake regulation and poor sleep continuity, both of which have been linked to depressive symptoms and future depression risk [22, 23, 37-39]. Thus, wearable-derived physical activity and sleep phenotypes may serve as interpretable digital behavioral markers for long-term MDD risk characterization in real-world settings.

The sequential model comparisons demonstrated the incremental value of longitudinal digital monitoring. Adding monthly wearable-derived features improved model performance beyond baseline wearable summaries, suggesting that repeated monitoring captured dynamic changes in behavior that were not reflected in baseline activity and sleep patterns alone. Other information domains also contributed complementary information: MDD PRS captured stable inherited liability, baseline wearable features represented habitual behavioral profiles before follow-up, and the selected wearable feature × MDD PRS interaction terms reflected behavioral associations that differed by genetic liability. Together, these results suggest that integrating genetic, baseline behavioral, and longitudinal digital information can provide a more complete characterization of incident MDD risk than any single information domain alone.

Our findings should be interpreted in the context of several limitations. First, requiring linked genomic, EHR, and Fitbit data, together with the bring-your-own-device nature of Fitbit contribution, may have selected for more health-conscious or technologically engaged participants, limiting generalizability to broader populations. Second, although we used definitions consistent with prior AoURP wearable studies [25, 26, 32, 40] and excluded participants with EHR-recorded MDD before or during the baseline wearable period, left censoring, diagnostic delay, residual confounding, and undiagnosed cases cannot be fully excluded. These are inherent limitations of EHR-based observational studies. Third, although 284 participants developed EHR-recorded incident MDD, statistical power may have been limited for interaction analyses and PRS-stratified estimates; replication in larger cohorts is warranted. Fourth, because of limited cross-ancestry PRS transferability and the smaller sample size of non-European ancestry groups, our analyses were restricted to participants of European ancestry; validation in more diverse populations is needed.

In conclusion, this study shows that MDD PRS and longitudinal wearable-derived behavioral features provide additional information for characterizing EHR-recorded incident MDD risk. The added value of monthly time-varying wearable features highlights the importance of long-term digital monitoring for capturing dynamic behavioral information beyond baseline activity and sleep patterns. The associations of sedentary time and sleep duration variability with incident MDD differed across MDD PRS levels, suggesting that some wearable-derived behavioral phenotypes may carry different risk information according to genetic liability. In exploratory analyses, comparable estimated risk corresponded to different behavioral levels across MDD PRS levels, with more favorable behavioral profiles observed among participants with higher MDD PRS. These findings support a genetically informed longitudinal digital phenotyping framework for MDD risk characterization, with potential implications for future personalized behavioral risk monitoring and prevention strategies.

## Method

### Data source

This study used data from the AoURP, an ongoing national longitudinal cohort study funded by the US National Institutes of Health [34]. Detailed descriptions of the study design, data collection, and data curation have been published previously [34, 41]. We used the AoURP controlled tier dataset, version 7 (C2022Q4R13), which included participants recruited between May 2018 and July 2022. The linked AoURP data infrastructure enabled integration of genome-wide genetic data, longitudinal Fitbit wearable data, EHR data, and participant-reported health survey data. In the present analysis, genomic data were used to derive MDD

PRS as a measure of genetic liability; Fitbit data were used to construct longitudinal wearable-derived physical activity and sleep features; EHR data were used to identify MDD diagnoses and define censoring; and survey data were used to define sociodemographic, lifestyle, and health-related covariates. Genomic data were generated from participant biospecimens processed through the All of Us Genome Centers and Biobank [34, 42]. For participants who consented to share EHR and Fitbit data, the research dataset included both historical records available from before AoURP enrollment and records accrued during follow-up, from participating health care provider organizations and linked Google Fitbit accounts, respectively [34, 41, 43].

**Ethical considerations**

This study was conducted as a secondary analysis of de-identified AoURP data by authorized researchers within the secure All of Us Researcher Workbench. The analysis involved only existing de-identified data and did not involve direct contact with participants; therefore, no additional institutional review board approval was required. All participants provided informed consent for research use of their data at enrollment in the AoURP.

**EHR-based identification of MDD diagnoses**

MDD diagnoses were identified from EHR condition records using standardized condition concept identifiers in the Observational Medical Outcomes Partnership (OMOP) Common Data Model. The OMOP Common Data Model harmonizes diagnoses across contributing health care systems and maps to source vocabularies including ICD-9-CM, ICD-10-CM, and SNOMED CT [44]. The OMOP concept identifiers and corresponding ICD-9-CM and ICD-10-CM codes used to identify MDD diagnoses are provided in Supplementary Table 1. EHR diagnosis dates were used to define prior MDD history and incident MDD events in the analytic cohort construction described below. These dates indicate when an MDD diagnosis was recorded in the EHR and were not interpreted as dates of symptom onset.

**Polygenic risk score calculation**

We used biallelic autosomal single-nucleotide variants (SNVs) that had passed AoURP initial genomic quality-control procedures [42]. Additional filtering removed duplicate-position variants and low-quality genotypes, with variant-level filters applied for missingness and Hardy–Weinberg equilibrium following the procedures described previously [31]. Because PRS performance has limited cross-ancestry transferability and the available sample size for non-European ancestry groups was limited, PRS calculation and subsequent analyses were restricted to participants of genetically inferred European ancestry.

We constructed an MDD PRS using the publicly available European-ancestry release of the Psychiatric Genomics Consortium (PGC) MDD2025 GWAS summary statistics, comprising 412,305 MDD cases and 1,588,397 controls [45]. The accompanying dataset and study lists did not report the inclusion of AoURP participants. Nonambiguous single-nucleotide polymorphisms (SNPs) with minor allele frequency greater than 0.01 were retained from the GWAS summary statistics. Linkage disequilibrium was modeled using the European ancestry reference panel from the 1000 Genomes Project Phase 3 [46]. Posterior SNP effect sizes were estimated using PRS-continuous shrinkage (PRS-CS), a Bayesian polygenic scoring method that applies continuous shrinkage priors to account for linkage disequilibrium, with the global shrinkage parameter auto-estimated [47].

Individual-level PRS values were calculated by summing effect-allele dosages weighted by the corresponding PRS-CS posterior effect sizes. To reduce residual confounding by population structure, raw PRS values were residualized on the top 10 genetic principal components derived from genome-wide genotype data and then standardized to a mean of 0 and standard deviation of 1. For readability, this standardized residualized score is referred to as the MDD PRS throughout the manuscript. To assess MDD PRS performance, we examined its association with any recorded MDD diagnosis among participants with available PRS and EHR data, irrespective of Fitbit linkage, using logistic regression adjusted for age and sex. This analysis was conducted separately from the primary incident MDD analysis.

**Fitbit physical activity and sleep data**

Fitbit provides daily summaries of steps and time spent in four proprietary activity categories: sedentary, lightly active, fairly active, and very active. These categories are informed by estimated metabolic equivalent of task (MET) values, with sedentary activity defined as <1.5 METs, lightly active as 1.5–3 METs, fairly active as 3–6 METs, and very active as ≥6 METs or high-cadence activity [48]. These Fitbit categories broadly correspond to conventional activity intensity definitions [49]. For readability, we refer to lightly active, fairly active, and very active minutes as light, moderate, and vigorous physical activity, respectively, throughout the manuscript. Daily steps were used as an interpretable measure of ambulatory activity volume [19, 50] and sedentary time and light, moderate, and vigorous physical activity were used to characterize the distribution of daily activity [51]. To ensure data quality, we followed prior AoURP wearable studies [25, 26] in defining valid activity days as days with at least 10 hours of wear time and a total daily step count between 100 and 45,000. Wear time was defined as any hour with recorded step data.

Fitbit also provides daily sleep summaries, including total sleep time (hereafter referred to as sleep duration), time in bed, and Fitbit-estimated sleep stages. Although Fitbit devices have shown reasonable performance in detecting sleep–wake states, validation studies have reported more limited accuracy for classifying sleep stages against gold-standard sleep assessment [52, 53]. Therefore, to prioritize more robust and interpretable sleep measures, we focused on sleep duration and sleep efficiency, calculated as the ratio of sleep duration to time in bed. For sleep data quality, we retained only sleep periods flagged by Fitbit as "main sleep", defined as the longest sleep period per day, to reduce the influence of naps [26].

**Baseline and time-varying wearable feature aggregation**

We followed prior AoURP studies to define the baseline wearable period and aggregate post-baseline wearable data into monthly time-varying features [25, 26, 31]. Daily Fitbit physical activity and sleep measures were aggregated into two temporal feature sets: baseline and monthly time-varying wearable features. The first 180 days of Fitbit monitoring were defined as the baseline wearable period, and valid daily records during this window were aggregated to represent each participant's habitual patterns before follow-up. After the baseline wearable period, valid daily Fitbit records were aggregated into monthly intervals to construct time-varying wearable features during follow-up. Months were retained only if they included at least 15 valid days for both physical activity and sleep. These monthly features were intended to capture longitudinal variation in daily-life behavior beyond each participant's baseline behavioral profile.

For both baseline and monthly time-varying wearable features, daily Fitbit measures were summarized across valid days within each aggregation interval. Daily values were averaged to construct features for daily steps, sedentary time, light physical activity, moderate physical activity, vigorous physical activity, sleep duration, and sleep efficiency. Sleep duration variability was calculated as the standard deviation of daily sleep duration within each aggregation interval.

**Analytic cohort construction and incident MDD definition**

Participants were eligible for inclusion if they had linked genomic, Fitbit wearable, and EHR data. Fitbit records collected before age 18 years were excluded. Participants were required to have a Fitbit monitoring span of at least 180 calendar days to define the baseline wearable period, with at least 15 valid days of both physical activity and sleep data during this period.

To focus on EHR-based incident MDD, we excluded participants with any recorded MDD diagnosis before Fitbit monitoring or during the 180-day baseline wearable period. We also excluded participants with any recorded diagnosis of bipolar disorder, schizophrenia, or schizoaffective disorder [40, 54], because these conditions may complicate the interpretation of EHR-recorded depressive diagnoses and behavioral risk patterns (Supplementary Table 1). Incident MDD was defined as the first EHR-recorded MDD diagnosis occurring after the baseline period, consistent with prior AoURP wearable studies [25, 26]. This approach was intended to reduce inclusion of participants with pre-existing MDD and to mitigate potential reverse causation from early depressive symptoms influencing baseline wearable-derived behavior [31].

## Statistical analysis

### *Descriptive analysis*

Participant characteristics, baseline physical activity features, and baseline sleep features were summarized across MDD PRS quartiles. Continuous variables were reported as medians and interquartile ranges, and categorical variables were reported as counts and percentages. Differences across PRS quartiles were assessed using Kruskal–Wallis tests for continuous variables and Pearson's $\chi^2$ tests for categorical variables.

### *Time-varying Cox proportional hazards models*

Time-varying Cox proportional hazards regression models were used to examine the associations among monthly wearable features, MDD PRS, and time to EHR-recorded incident MDD. The time-to-event outcome was defined as the time from the end of the 180-day baseline wearable period to the first EHR-recorded incident MDD diagnosis. For the time-varying Cox models, follow-up time was measured from the index date and represented as monthly person-time records aligned with the monthly wearable feature aggregation windows. Only records with valid monthly wearable features contributed to the analyses, and missing monthly wearable features were not interpolated or imputed. Participants without incident MDD were censored at the earliest of their last clinical encounter (most recent EHR record from laboratory, diagnosis, procedure, or vital sign domains) or the administrative censoring date of 1 July 2022.

For each wearable feature, we fitted a separate model including the monthly time-varying wearable feature of interest, MDD PRS, the wearable feature × MDD PRS interaction term, age, sex, BMI, education, smoking status, alcohol use, and the corresponding baseline wearable feature. The wearable feature × MDD PRS interaction term was used to evaluate whether genetic liability modified the association between wearable-derived behavior and incident MDD risk. Including the corresponding baseline wearable feature helped account for habitual behavioral patterns, allowing the monthly time-varying feature to capture dynamic behavioral information beyond each participant's baseline profile over time. Continuous variables were first modeled using restricted cubic splines with three knots. Spline terms were retained when there was statistically significant evidence of nonlinearity; otherwise, variables were modeled linearly. To facilitate comparison across wearable features with different units, we reported the HR for each wearable feature by comparing the 75th with the 25th percentile. For a given feature, an HR greater than 1 indicated a higher model-estimated risk of EHR-recorded incident MDD at the 75th percentile than at the 25th percentile, whereas an HR less than 1 indicated a lower model-estimated risk. Missing values in baseline covariates were handled using bootstrap-based multiple imputation with predictive mean matching. The proportional hazards assumption was tested using scaled Schoenfeld residuals.

*PRS-stratified behavioral levels at comparable estimated risk*

To examine whether comparable estimated risk of incident MDD corresponded to different behavioral levels across genetic liability, we estimated PRS-stratified behavioral levels from the fitted Cox models. For each wearable feature, model-predicted HRs were calculated across the evaluated range of feature values at selected PRS percentiles. The risk-equivalent behavioral level was defined as the feature value at which the predicted HR equaled 1.00, indicating no higher or lower model-estimated risk on the HR scale. When the predicted HR curve did not cross 1.00 within the evaluated range, the behavioral level was considered not estimable. These estimates were interpreted as model-derived summaries of behavioral levels corresponding to comparable estimated risk across PRS strata, rather than causal thresholds or clinical intervention targets.

*Incremental model comparison for MDD risk characterization*

To evaluate the incremental information provided by genetic and wearable-derived measures for EHR-recorded incident MDD risk characterization, we compared a series of sequentially nested Cox models. Model 1 included base covariates: age, sex, BMI, education, smoking status, and alcohol use. Model 2 additionally included MDD PRS to evaluate the added information from genetic liability. Model 3 further added baseline wearable features to evaluate the contribution of habitual physical activity and sleep patterns. Model 4 added monthly time-varying wearable features to assess the additional information provided by longitudinal wearable monitoring during follow-up. Model 5 added the statistically significant wearable feature × MDD PRS interaction terms identified in the primary Cox models to characterize the incremental contribution of PRS-stratified behavioral associations. For each sequential model extension, changes in Akaike information criterion (ΔAIC) and Harrell's C-index (ΔC-index) relative to the preceding model were calculated. Likelihood ratio tests were used to assess statistical evidence for improved model fit between nested models.

*Sensitivity analysis*

Because an EHR-recorded diagnosis may occur after the onset of depressive symptoms, wearable-derived behavioral changes shortly before diagnosis may reflect emerging illness rather than risk patterns. To assess the robustness of the primary findings to potential reverse causation and diagnostic delay, we repeated the primary Cox analyses after excluding wearable observations within 6 months before the first EHR-recorded incident MDD diagnosis among participants who developed MDD. This analysis evaluated whether the observed associations were driven by behavioral changes occurring close to the recorded diagnosis date.

**Data availability**

To protect participant privacy, the data used in this study are available to approved researchers through the All of Us Research Workbench (https://workbench.researchallofus.org/login) following registration, completion of required ethics training, and attestation to the data use agreement.

**Code availability**

Code used for this study is available to approved researchers within the All of Us Research Workbench platform upon request to the corresponding authors.

**Acknowledgements**

The authors gratefully acknowledge the All of Us participants for their contributions, without whom this research would not have been possible. They also thank the National Institutes of Health's All of Us Research Program for making available the participant data examined in this study. R.J.B.D. is supported by the following: (1) National Institute for Health and Care Research (NIHR) Biomedical Research Centre (BRC) at South London and Maudsley National Health Service (NHS) Foundation Trust and King's College London; (2) Health Data Research UK, which is funded by the UK Medical Research Council, Engineering and Physical Sciences Research Council, Economic and Social Research Council, Department of Health and Social Care (England), Chief Scientist Office of the Scottish Government Health and Social Care Directorates, Health and Social Care Research and Development Division (Welsh government), Public Health Agency (Northern Ireland), British Heart Foundation, and Wellcome Trust; (3) the BigData@Heart consortium, funded by the Innovative Medicines Initiative 2 Joint Undertaking (which receives support from the European Union's Horizon 2020 research and innovation program and the European Federation of Pharmaceutical Industries and Associations, in collaboration with 20 academic and industry partners and the European Society of Cardiology); (4) the NIHR University College London Hospitals BRC; (5) the NIHR BRC at South London and Maudsley (related to attendance at the American Medical Informatics Association) NHS Foundation Trust and King's College London; (6) the UK Research and Innovation London Medical Imaging & Artificial Intelligence Centre for Value-Based Healthcare; (7) the NIHR Applied Research Collaboration South London at King's College Hospital NHS Foundation Trust; and (8) the Wellcome Trust. The funders had no role in the design and conduct of the study; collection, management, analysis, and interpretation of the data; preparation, review, and approval of the manuscript; or decision to submit the manuscript for publication.

**Author contributions**

Y.Z. had full access to all study data and takes responsibility for the integrity of the data and the accuracy of the data analysis. Y.Z., A.A.F., and R.J.B.D. contributed to the study concept and design. Y.Z. conducted the statistical analysis and drafted the manuscript. Y.Z., A.A.F., R.Z., H.K., S.S., C.S., and R.J.B.D. contributed to interpretation of the findings. R.J.B.D. and A.A.F. obtained funding. All authors critically reviewed the manuscript for important intellectual content and approved the final version of the manuscript.

**Competing interests**

A.A.F. and R.J.B.D. are cofounders of Onsentia. A.A.F. holds shares in Google. All other authors declare no other conflicts of interest.

**Table 1. Baseline characteristics of the analytic cohort stratified by major depressive disorder polygenic risk score (MDD PRS) quartile.**

| Characteristic | Overall (N=3030) | Q1 PRS (N=758) | Q2 PRS (N=757) | Q3 PRS (N=757) | Q4 PRS (N=758) | P value |
|---|---|---|---|---|---|---|
| Age, years | 52.4 [36.9, 62.7] | 53.9 [37.1, 63.6] | 51.8 [36.1, 62.5] | 51.5 [37.5, 62.5] | 51.8 [36.9, 62.0] | 0.516 |
| Female, n/N (%) | 2113/2959 (71.4%) | 541/745 (72.6%) | 512/740 (69.2%) | 531/740 (71.8%) | 529/734 (72.1%) | 0.472 |
| BMI, kg/m² | 27.6 [24.2, 32.2] | 26.8 [23.9, 31.2] | 27.9 [24.2, 33.1] | 27.1 [24.2, 31.7] | 28.4 [24.8, 33.0] | <0.001 |
| Education, n/N (%) | | | | | | <0.001 |
| College degree | 2166/2944 (73.6%) | 597/743 (80.3%) | 558/735 (75.9%) | 535/739 (72.4%) | 476/727 (65.5%) | |
| Some college | 623/2944 (21.2%) | 119/743 (16.0%) | 137/735 (18.6%) | 175/739 (23.7%) | 192/727 (26.4%) | |
| No college | 155/2944 (5.3%) | 27/743 (3.6%) | 40/735 (5.4%) | 29/739 (3.9%) | 59/727 (8.1%) | |
| Lifetime alcohol use ⩾1 drink, n/N (%) | 2956/3017 (98.0%) | 735/755 (97.4%) | 740/753 (98.3%) | 738/754 (97.9%) | 743/755 (98.4%) | 0.461 |
| Lifetime smoking ⩾ 100 cigarettes, n/N (%) | 963/2979 (32.3%) | 190/745 (25.5%) | 257/744 (34.5%) | 228/745 (30.6%) | 288/745 (38.7%) | <0.001 |
| Fitbit monitoring duration, years | 5.05 [3.08, 6.62] | 4.95 [3.02, 6.64] | 5.07 [3.17, 6.68] | 5.03 [3.00, 6.55] | 5.07 [3.18, 6.59] | 0.807 |
| Baseline daily steps | 8394 [6342, 10593] | 8457 [6531, 10441] | 8711 [6718, 10793] | 8267 [6149, 10599] | 8218 [6014, 10465] | 0.007 |
| Baseline sedentary minutes/day | 748 [680, 845] | 748 [676, 849] | 746 [685, 837] | 754 [681, 853] | 746 [682, 844] | 0.928 |
| Baseline light PA minutes/day | 224.5 [180.9, 269.4] | 227.5 [183.9, 267.8] | 222.5 [181.0, 265.9] | 225.5 [180.2, 272.1] | 222.7 [179.2, 269.6] | 0.846 |
| Baseline moderate PA minutes/day | 17.7 [9.2, 36.4] | 17.6 [9.5, 35.3] | 18.4 [10.4, 38.6] | 17.3 [8.8, 36.1] | 17.8 [8.2, 36.8] | 0.117 |
| Baseline vigorous PA minutes/day | 16.5 [7.4, 30.3] | 17.6 [8.8, 29.5] | 18.0 [8.5, 31.4] | 14.9 [6.8, 31.2] | 15.4 [6.1, 28.4] | <0.001 |
| Baseline sleep duration, h | 6.83 [6.20, 7.35] | 6.85 [6.21, 7.36] | 6.83 [6.22, 7.37] | 6.84 [6.21, 7.34] | 6.78 [6.17, 7.33] | 0.791 |
| Baseline sleep duration variability, h | 1.42 [1.12, 1.81] | 1.36 [1.09, 1.73] | 1.41 [1.12, 1.80] | 1.43 [1.13, 1.82] | 1.50 [1.16, 1.86] | <0.001 |
| Baseline sleep efficiency, % | 90.3 [87.8, 93.5] | 90.6 [88.0, 93.7] | 90.5 [88.0, 93.4] | 90.1 [87.9, 93.5] | 90.1 [87.5, 93.2] | 0.075 |

Note. Q1 to Q4 represent ascending quartiles of MDD PRS, with Q1 indicating the lowest and Q4 the highest genetic liability. Continuous variables are presented as median [interquartile range], and categorical variables as n/N (%). BMI, body mass index. Education was categorized as college degree (college graduate or advanced degree), some college (one to three years of college), or no college (high school diploma, GED, or less). PA, physical activity.

**Table 2. Associations between time-varying wearable features and incident major depressive disorder risk.**

| Feature | 75th vs 25th percentile contrast | HR (95% CI) | P value |
|---|---|---|---|
| **Physical activity** | | | |
| Daily steps | 10.87 vs 5.95 thousand steps/day | 0.59 (0.46–0.77) | <0.001 |
| Sedentary time | 776.37 vs 641.86 min/day | 1.09 (0.94–1.26) | 0.248 |
| Light physical activity | 274.33 vs 186.24 min/day | 0.66 (0.54–0.81) | <0.001 |
| Moderate physical activity | 27.11 vs 7.56 min/day | 0.98 (0.85–1.12) | 0.738 |
| Vigorous physical activity | 34.42 vs 6.31 min/day | 0.78 (0.62–0.97) | 0.027 |
| **Sleep** | | | |
| Sleep duration variability | 1.56 vs 0.90 h | 1.29 (1.12–1.48) | <0.001 |
| Sleep efficiency | 92% vs 87% | 0.90 (0.81–0.99) | 0.039 |
| Sleep duration | 7.30 vs 6.19 h | 1.02 (0.89–1.17) | 0.783 |

Note: Hazard ratios were estimated from separate time-varying Cox proportional hazards models for each wearable feature and compare the 75th with the 25th percentile of the corresponding monthly time-varying feature. Models were adjusted for age, sex, body mass index, education, smoking status, alcohol use, MDD PRS, and the corresponding baseline wearable feature. Full estimates from the physical activity and sleep models are provided in Supplementary Tables 3 and 4, respectively. Sleep duration showed evidence of nonlinearity; the reported 75th versus 25th percentile contrast is provided for consistency and does not fully characterize the association. HR, hazard ratio; CI, confidence interval; MDD PRS, major depressive disorder polygenic risk score.

**Table 3. PRS-stratified associations between wearable features and incident major depressive disorder for features with significant wearable feature × MDD PRS interactions.**

| Feature | MDD PRS level | 75th vs 25th percentile contrast | HR (95% CI) | P value |
|---|---|---|---|---|
| Sedentary time | 25th percentile | 776.37 vs 641.86 min/day | 0.98 (0.82–1.17) | 0.822 |
| | 50th percentile | 776.37 vs 641.86 min/day | 1.09 (0.95–1.26) | 0.225 |
| | 75th percentile | 776.37 vs 641.86 min/day | 1.22 (1.05–1.41) | 0.009 |
| Sleep duration variability | 25th percentile | 1.56 vs 0.90 h | 1.18 (0.98–1.42) | 0.077 |
| | 50th percentile | 1.56 vs 0.90 h | 1.29 (1.13–1.49) | <0.001 |
| | 75th percentile | 1.56 vs 0.90 h | 1.42 (1.24–1.62) | <0.001 |

Note: PRS-stratified associations were estimated only for wearable features with statistically significant wearable feature × MDD PRS interaction terms in the primary time-varying Cox models. Hazard ratios compare the 75th with the 25th percentile of each monthly time-varying wearable feature at the specified MDD PRS percentile (25th, 50th, and 75th). Models were adjusted for age, sex, body mass index, education, smoking status, alcohol use, MDD PRS, and the corresponding baseline wearable feature. P values correspond to the PRS-stratified hazard ratio estimates. CI, confidence interval; HR, hazard ratio; MDD PRS, major depressive disorder polygenic risk score.

**Table 4. Incremental contribution of MDD PRS and wearable-derived features to incident MDD risk models.**

| Model | Predictors added | AIC | ΔAIC | C-index | ΔC-index | LR $\chi^2$ (P value) |
|---|---|---|---|---|---|---|
| Model 1: Base covariates | Age, sex, BMI, education, smoking status, and alcohol use | 4222.66 | — | 0.637 | — | — |
| Model 2: + MDD PRS | MDD PRS | 4209.32 | −13.35 | 0.650 | 0.013 | 15.25 ($P < 0.001$) |
| Model 3: + Baseline wearable features | Baseline wearable-derived features | 4183.33 | −25.98 | 0.679 | 0.029 | 41.18 ($P < 0.001$) |
| Model 4: + Time-varying wearable features | Monthly time-varying wearable-derived features | 4162.15 | −21.18 | 0.703 | 0.024 | 38.79 ($P < 0.001$) |
| Model 5: + Selected wearable × MDD PRS interactions | Sedentary time × MDD PRS and sleep duration variability × MDD PRS | 4156.18 | −5.98 | 0.705 | 0.002 | 10.15 ($P = 0.006$) |

Note: Models were sequentially nested, and the "Predictors added" column lists variables introduced relative to the preceding model. ΔAIC and ΔC-index were calculated relative to the preceding model. LR $\chi^2$ statistics and P values were obtained from likelihood ratio tests comparing each model with the preceding nested model. Lower AIC and higher C-index indicate better model fit and discrimination, respectively. AIC, Akaike information criterion; BMI, body mass index; LR, likelihood ratio; MDD, major depressive disorder; PRS, polygenic risk score.

**Figure 1. Study design and analytic framework.** EHR, electronic health record; MDD, major depressive disorder; PA, physical activity; PRS, polygenic risk score.

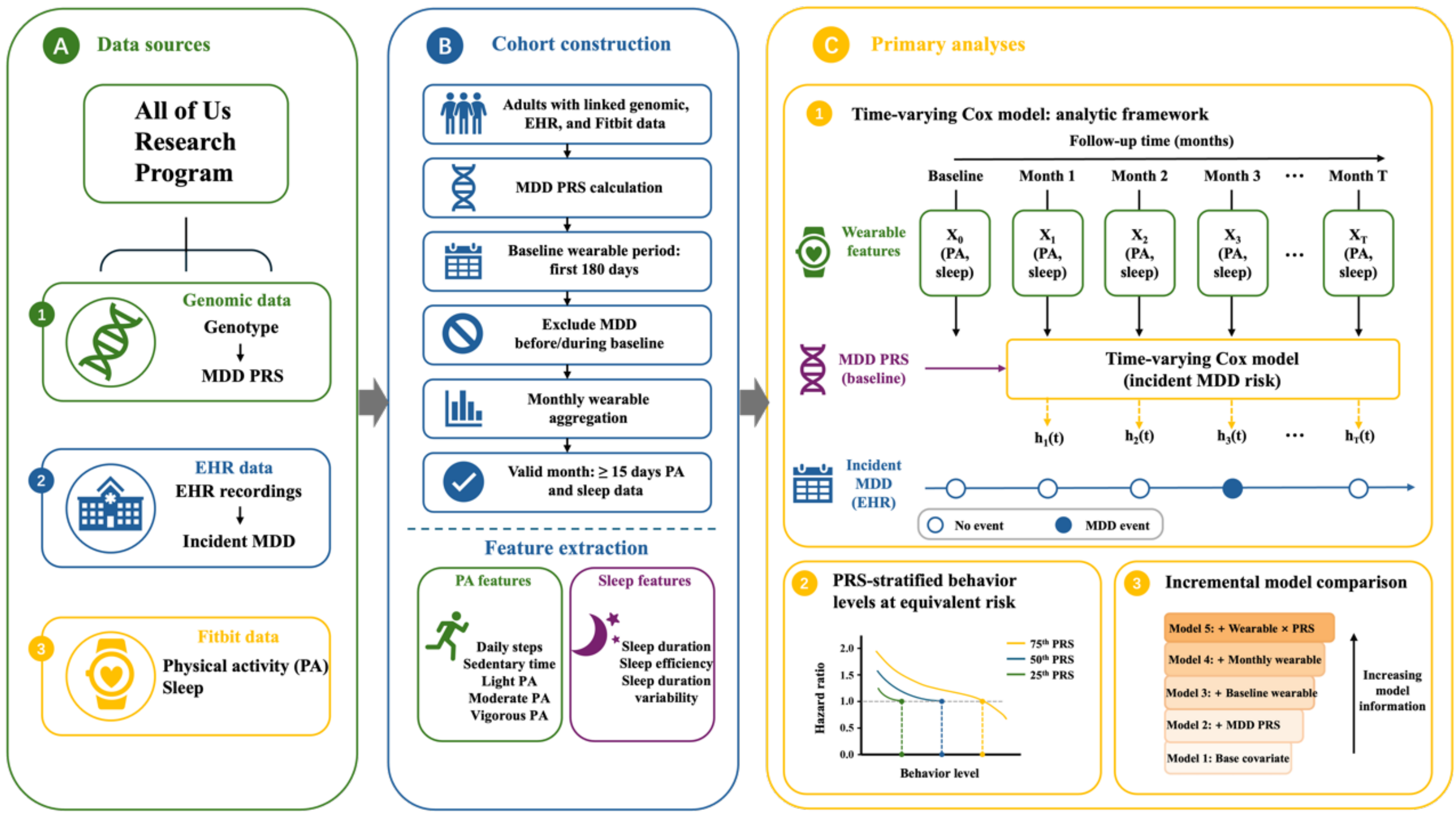

**Figure 2. Wearable feature–hazard ratio curves for incident MDD stratified by MDD PRS percentile.** Model-estimated hazard ratios are shown across the observed range of (a) daily steps, (b) sedentary time, (c) sleep duration variability, and (d) sleep duration at the 25th, 50th, and 75th percentiles of MDD PRS. Shaded areas indicate 95% confidence intervals. The dashed horizontal line indicates hazard ratio = 1.00, corresponding to neither higher nor lower model-estimated risk on the HR scale. MDD, major depressive disorder; PRS, polygenic risk score.

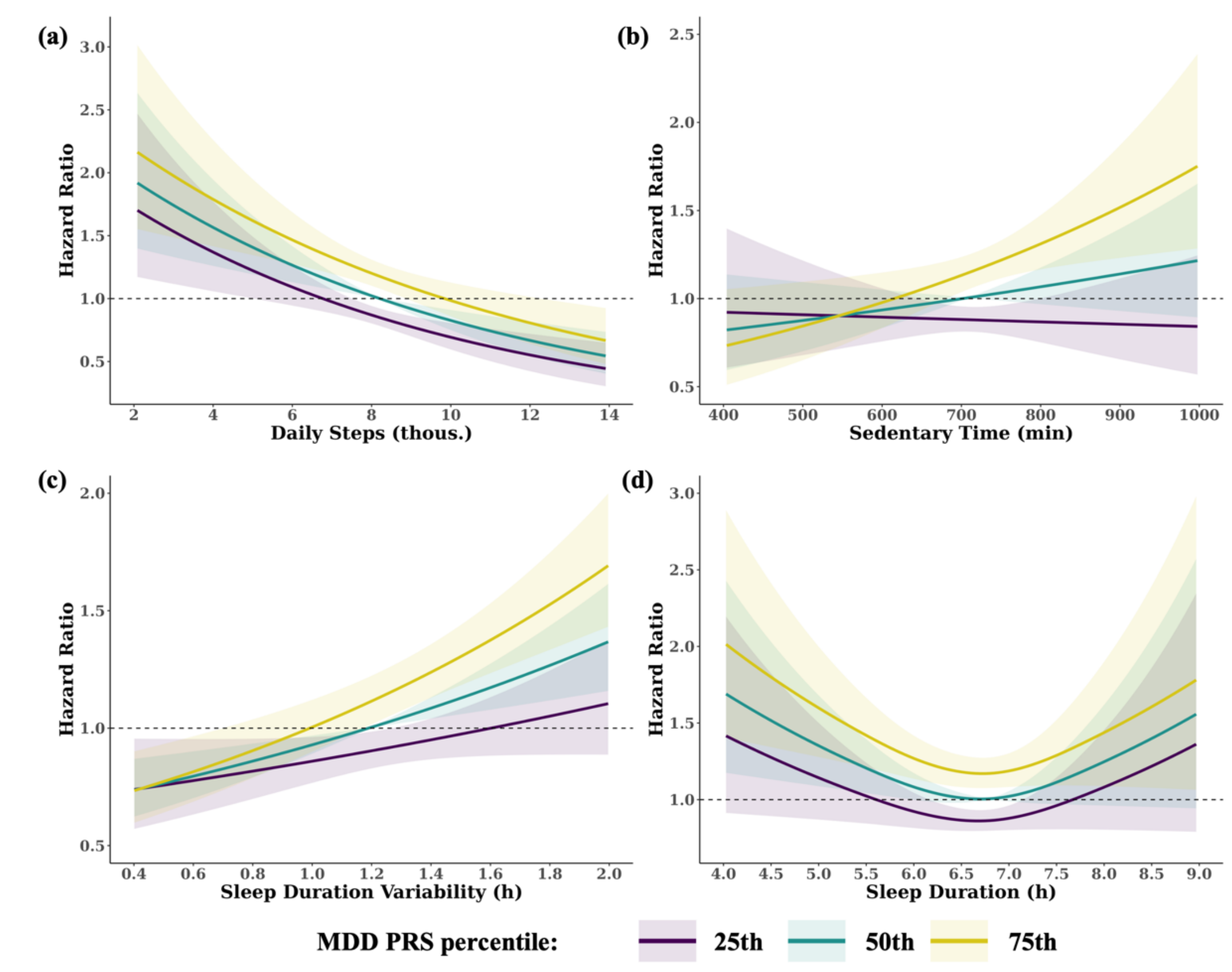